\documentclass{article}
\usepackage{spconf,amsmath,graphicx}
\usepackage{subfigure}
\usepackage{multirow}
\usepackage{booktabs}
\usepackage{amsmath, amssymb, mathtools}
\usepackage{enumitem}
\usepackage{url}
\usepackage{hyperref}
\usepackage{url}
\usepackage[symbol]{footmisc}

\title{InferGrad: Improving Diffusion Models for Vocoder by Considering Inference in Training}
\name{Zehua Chen$^{1,2}$\thanks{Work done during an internship at Microsoft.}, Xu Tan$^{3}$, Ke Wang$^{2}$, Shifeng Pan$^{2}$, Danilo Mandic$^{1}$, Lei He$^{2}$\thanks{Lei He is the corresponding author.}, Sheng Zhao$^{2}$}
\address{$^1$Department of EEE, Imperial College London, SW7 2AZ, London, UK\\
  $^2$Microsoft Azure Speech, 100080, Beijing, China\\
  $^3$Microsoft Research Asia, 100080, Beijing, China}
\begin{document}
\ninept
\maketitle
\begin{abstract}
Denoising diffusion probabilistic models (diffusion models for short) require a large number of iterations in inference to achieve the generation quality that matches or surpasses the state-of-the-art generative models, which invariably results in slow inference speed. Previous approaches aim to optimize the choice of inference schedule over a few iterations to speed up inference. However, this results in reduced generation quality, mainly because the inference process is optimized separately, without jointly optimizing with the training process. In this paper, we propose InferGrad, a diffusion model for vocoder that incorporates inference process into training, to reduce the inference iterations while maintaining high generation quality. More specifically, during training, we generate data from random noise through a reverse process under inference schedules with a few iterations, and impose a loss to minimize the gap between the generated and ground-truth data samples. Then, unlike existing approaches, the training of InferGrad considers the inference process. The advantages of InferGrad are demonstrated through experiments on the LJSpeech dataset showing that InferGrad achieves better voice quality than the baseline WaveGrad under same conditions while maintaining the same voice quality as the baseline but with $3$x speedup ($2$ iterations for InferGrad vs $6$ iterations for WaveGrad).

\end{abstract}
\vspace{-0.15cm}
\section{Introduction}
\vspace{-0.15cm}
\label{sec:intro}
Deep generative models have demonstrated supreme ability in high-fidelity speech synthesis~\cite{WaveNet,HiFi-GAN,WaveGrad,tan2021survey}. The type of these models mainly includes autoregressive models \cite{WaveNet,SampleRNN,EfficientNeuralAudioSynthesis}, normalizing flows \cite{NINF,ParallelWaveNet,Glow}, variational autoencoders \cite{VAE} and generative adversarial networks \cite{HiFi-GAN,MelGAN,GAN-TTS}. Recently, denoising diffusion probabilistic models (DDPMs, diffusion models for short) are emerging \cite{Songyang1,DDPM,Songyang2,VariationalDiffusion}. They consist of two processes: 1) diffusion/forward process, where the data distribution is transformed into a known prior noise distribution, e.g., the Gaussian noise; 2) denoising/reverse process, where the data samples are gradually recovered from the random noise with the learned score function. Diffusion models have been extensively used for generation tasks \cite{WaveGrad,DiffWave,Grad-TTS,PriorGrad,LSGM,BDDM} since their capability of generating high-fidelity samples matches the state-of-the-art autoregressive and GANs based methods.

In diffusion models, when the number of steps in reverse process matches that in forward process, the learned gradient information is fully utilized and the generation quality is maximized. However, a large number of reverse steps decreases the inference speed, which limits the application scenarios of diffusion models. Thus, previous works proposed to reduce the number of reverse steps, which can accelerate the generation process, but at the cost of reduced generation quality. Many of these works explored to optimize the choice of inference noise schedule (inference schedule for short), without changing the training process. In WaveGrad \cite{WaveGrad}, grid search was employed to find the optimal inference schedule, while in DiffWave \cite{DiffWave} it was manually defined. In \cite{noiseestimationDiffusion}, they trained an additional network to estimate the noise level and then used a rule-based module to update the inference schedule parameters at each inference step. In BDDMs \cite{BDDM}, they also trained an extra scheduling network that was used to optimize the choice of inference schedule. In DDIM \cite{DDIM}, the diffusion process was reparameterized as a non-Markovian process, which supports the inference process with a subset of the training noise schedule. For these methods, the advantage is that they do not require retraining the DDPMs, which keeps the flexibility of DDPMs to the choice of inference schedule. However, as the noise schedule in training process and inference process are decoupled, they are not able to improve the generation quality of the selected inference schedule at training stage. Some research works considered to change the training process. The Grad-TTS \cite{Grad-TTS} and the PriorGrad \cite{PriorGrad} transformed the data distribution into a data-driven prior noise distribution which is obtained from the conditioning information. The LSGM \cite{LSGM} trained diffusion models in a small latent space by using the variational autoencoder framework. In \cite{DiffusionBeatsGANs}, they additionally trained a classifier on the noisy data samples and then used the gradients of this classifier to guide the sampling. These methods are innovative in utilizing the conditioning information or the latent space. Being different from these methods, we choose to incorporate the information of inference schedule into training and optimize the DDPMs for the considered inference schedules. 

\begin{figure*}[htbp]
\centerline{\includegraphics[width=\linewidth]{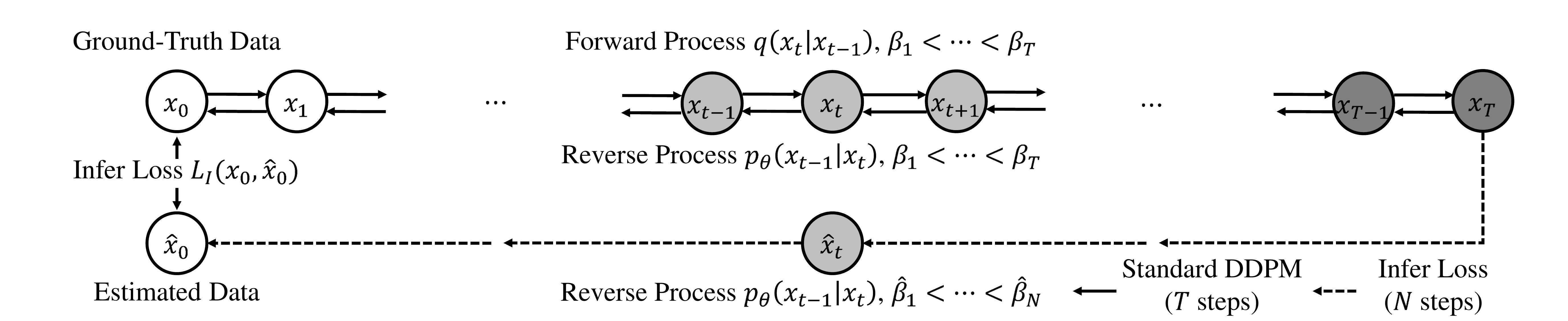}}
\vspace{-0.2cm}
\caption{Overview of InferGrad design.}
\label{fig:flowchart}
\vspace{-0.2cm}
\end{figure*}

\vspace{-0.08cm}
In this work, we propose InferGrad, a diffusion model for vocoder that incorporates inference/reverse process into training, to reduce the inference iterations while maintaining high generation quality. After determining the range of inference schedules under a few reverse iterations, we generate the waveform from random noise following these inference schedules, and add an auxiliary loss in training objective to minimize the gap between generated and ground-truth waveform, where the loss contains both amplitude and phase information of the waveform after Fourier Transform. In vocoder task, the experiment results on LJSpeech dataset \cite{ljspeech17} show that 1) With a small number of inference steps, e.g., $2$, $3$ and $6$, InferGrad can effectively improve the audio sample quality over the WaveGrad baseline measured by both objective and subjective metrics; 2) InferGrad is more robust than baseline WaveGrad to inference schedules under a few iterations, which lowers the requirements on searching the optimal inference schedule. Audio samples are available at: \url{https://infergrad.github.io/}.

\section{Diffusion Probabilistic Model}

\vspace{-0.05cm}
Diffusion models consist of two processes: the forward process and the reverse process. The forward process is a Markov chain which injects Gaussian noise $\epsilon\sim\mathcal N(0,I)$ into data samples $x_{0}$ according to a predefined noise schedule $\beta$ with $ 0 < \beta_{1} < \dots < \beta_{T} < 1$. At each time step $t\in [1,\dots,T]$, the transition probability is:
\begin{align}
q(x_{t}|x_{t-1})=\mathcal{N}(x_{t};\sqrt{1-\beta_{t}}x_{t-1},\beta_{t}I).
\end{align} 

With the nice property of DDPMs and Gaussian noise \cite{DDPM}, we can directly obtain the noisy data distribution $q(x_{t}|x_{0})$ from $x_{0}$ by: $q(x_{t}|x_{0})=\mathcal  N(x_{t};\sqrt{\bar{\alpha}_{t}}x_{0},(1-\bar{\alpha}_{t})\epsilon)$, where $\alpha_{t}:=1-\beta_{t}$, and $\bar{\alpha}_{t}:=\prod_{s=1}^{t}\alpha_{s}$ denotes the corresponding noise level at time step $t$.

The reverse process is a denoising process which gradually removes the noise from $p(x_{T})\sim\mathcal N(0,I)$ to recover data $x_{0}$: 
\begin{gather}
p_{\theta}(x_{0},\cdots,x_{T-1}|x_{T})=\prod_{t=1}^{T}p_{\theta}(x_{t-1}|x_{t}),
\end{gather}
where the transition probability of each reverse step is parameterized as $p_{\theta}(x_{t-1}|x_{t})=\mathcal{N}(x_{t-1},\mu_{\theta}(x_{t},t),\sigma_{\theta}^{2}I)$. The variance $\sigma_{\theta}^{2}$ is usually predefined as $\frac{1-\bar{\alpha}_{t-1}}{1-\bar{\alpha}_{t}}\beta_{t}$ or $\beta_{t}$. One way to define the mean $\mu_{\theta}$ is \cite{DDPM,VariationalDiffusion}: $\mu_{\theta}(x_{t},t)=1/{\sqrt{\alpha_{t}}}(x_{t}-\beta_{t}/{\sqrt{1-\bar{\alpha}_{t}}}\epsilon_{\theta}(x_{t},t))$, where $\epsilon_{\theta}(x_{t},t))$ is the neural network estimated noise. The model is trained by maximizing the variation lower bound of the likelihood $p_{\theta}(x_{0})$. With the parameterization of $\mu_{\theta}$, in practice the training objective is usually defined as \cite{WaveGrad,DDPM,DiffWave}:
\begin{align}
\label{trainingobjective}
L_{D}(\theta)=\mathbb{E}_{x_{0},\epsilon,t}\left \| \epsilon - \epsilon_{\theta}(\sqrt{\bar{\alpha}_{t}}x_{0}+\sqrt{1-\bar{\alpha}_{t}}\epsilon,t) \right\|^2_{2}.
\end{align}

\section{InferGrad}
One important feature of DDPMs is that the noise schedule $\hat{\beta}$ used in reverse process can be different with the $\beta$ defined in forward process. With the number of diffusion steps $T$ and the number of reverse steps $N$, in order to improve the inference speed, DDPMs are usually employed with $N\ll T$. When $N$ is small, the choice of $\hat{\beta}$ becomes important for sample quality \cite{BDDM}. However, the training objective described in Equation (\ref{trainingobjective}) is designed to maximize the generalization ability of DDPMs for different inference schedules, rather than optimizing DDPMs with specified $\hat{\beta}$. In this work, aiming at improving the DDPMs sample quality when $N$ is small, we present InferGrad which improves DDPMs by incorporating a range of inference schedules that uses a few inference iterations into the training process. In this section, we first introduce our proposed infer loss that minimizes the gap between ground-truth waveform and the waveform generated from random noise following the incorporated inference schedules, and then discuss the principles to choose the inference schedules in our infer loss.

\subsection{Infer Loss}
As shown in Figure \ref{fig:flowchart}, given $N=T$, the DDPMs generate data samples $\hat{x}_{0}$ from $x_{T}$ as a mirror of the diffusion process. However, when we set $N\ll T$ for fast sampling, the distance between two adjacent inference steps is enlarged and the generation quality decreases. In InferGrad, we propose to use an infer loss $L_{I}$ to measure the distance between the ground truth $x_{0}$ and the generated sample $\hat{x}_{0}$. We incorporate the $L_{I}$ into the training objective of DDPMs as: 
\begin{align}
\label{newtrainingobjective}
L=L_{D}(\theta)+\lambda L_{I}(x_{0},\hat{x}_{0}),
\end{align}
where $L_{D}$ is defined in Equation (\ref{trainingobjective}) and $\lambda$ denotes the weight of infer loss. There are several considerations in designing $L_{I}$: 1) We measure the distance by using the ground truth data $x_{0}$ instead of the intermediate latent representation $x_{t}$ since it is not easy to accurately define the ground truth $x_{t}$ at each time step $t$. 2) The data $\hat{x}_{0}$ is generated from Gaussian noise $x_{T}$ with inference schedule $\hat{\beta}$. In this way, the entire reverse process is incorporated into optimization, i.e., reducing $L_{I}$ means improving the sample quality of DDPMs.      

The sample quality metrics used in $L_{I}$ should be close to human perception. Here we propose to use the multi-resolution short-time Fourier transform (STFT) loss function \cite{ParallelWaveGAN} as $L_{I}=\frac{1}{M}\sum_{m=1}^{M}L_{s}^{(m)}$, where $L_{s}$ is the single STFT loss and $M$ is the number of resolution. Different from \cite{ParallelWaveGAN}, we use both magnitude and phase information to calculate the loss term $L_{s}^{(m)}$, which can improve the voice quality. Consequently, the $L_{s}$ is computed between $x_{0}$ and $\hat{x}_{0}$ as follows:
\begin{align}
\label{newsingletrainingobjective}
L_{s}=\mathbb{E}_{x_{0},\hat{x}_{0}}[L_{mag}(x_{0},\hat{x}_{0})+L_{pha}(x_{0},\hat{x}_{0})],
\end{align}
where $L_{mag}$ is the $L_{1}$ loss of mel-scaled log STFT magnitude spectrum, and $L_{pha}$ is the $L_{2}$ loss of STFT phase spectrum.

\vspace{-0.15cm}
\subsection{Inference Schedules}
\label{scheduleprior}
When the number of inference steps $N$ is small, the DDPMs sample quality is sensitive to inference schedule $\hat{\beta}$. In InferGrad, we propose to include a range of $\hat{\beta}$ into $L_{I}$ thus enhancing the model robustness to inference schedules. Appropriately determining this range is helpful to keep the training stability and improve the generation towards inference. To our knowledge, there was no general guidance for manually choosing the reasonable range of optimal $\hat{\beta}$ in previous work. Here, for the vocoder task, we present several intuitive principles to determine the inference schedules $\hat{\beta}$ given $N\ll T$. 

Assuming the diffusion process destroys the data into nearly Gaussian noise $\mathcal N(0,I)$ with $ 0 < \beta_{1} < \dots < \beta_{T} < 1$ (schedule used in training). Following \cite{WaveGrad}, we denote $\bar{\alpha}_{t}=\prod_{s=1}^{t}(1-\beta_{s})$ as the noise level which ranges from $1$ at $t=0$ to nearly $0$ at $t=T$. Based on the schedules $\beta$ and $\bar{\alpha}$ used in training, we present several suggestions in determining the schedules $\hat{\beta}$ and $\hat{\bar{\alpha}}$ used in inference:

\begin{itemize}[leftmargin=*]
\item The range of $\hat{\beta}$. The $\hat{\beta}$ should follow $\beta_{1} \leq \hat{\beta}_{1} < \dots < \hat{\beta}_{N} < 1$, as the $\beta_{1}$ denotes the minimum noise scale used in training. When $N\ll T$, $\hat{\beta}_{N}$ is usually larger than $\beta_{T}$ to ensure $\hat{\bar{\alpha}}_{N}$ is located a in appropriate range. The meaning of $\hat{\bar{\alpha}}_{N}$ is discussed below. 
\vspace{-0.1cm}
\item The ratio between $\hat{\beta}_n$ and $\hat{\beta}_{n-1}$. When deciding $\hat{\beta}$, we recommend to start from $\hat{\beta}_{N}$ to $\hat{\beta}_{1}$. In order to ensure the denoising ability of DDPMs, the ratio between $\hat{\beta}_{n}$ and $\hat{\beta}_{n-1}$ should not be too large. In our test, it is not recommended to set $\frac{\hat{\beta}_{n}}{\hat{\beta}_{n-1}} > 10^{3}$.
\vspace{-0.1cm}
\item The value of $\hat{\bar{\alpha}}_{N}$. When starting reverse process from $\mathcal N(0,I)$, the $\hat{\bar{\alpha}}_{N}  \in [\bar{\alpha}_{T}, 1)$ should not be very close to $1$. The DDPMs should reach up to $\hat{\bar{\alpha}}_{N}$ from $\bar{\alpha}_{T}\approx 0$ in the first reverse step. Hence, when $\hat{\bar{\alpha}}_{N}$ is close to $1$, the denoising ability of DDPMs may not meet the requirement, which will result in distortions in samples. A value less than $0.7$ is recommended to use.    
\vspace{-0.05cm}
\end{itemize}
\vspace{-0.1cm}
Considering a range of $\hat{\beta}$ instead of a specified $\hat{\beta}$ in training makes model more robust to the choice of $\hat{\beta}$ in inference. If the optimal $\hat{\beta}$ in this range still needs to be found, grid search is one method to use.

\vspace{-0.2cm}
\section{Experiments}
\vspace{-0.05cm}
\subsection{Training Setup}
\textbf{Data}: We use the open-source $22.05$ kHz LJSpeech dataset \cite{ljspeech17} which contains around $24$ hours, $13,100$ audio samples from a female speaker. Except for LJ-$001$ and LJ-$002$,  $12577$ remaining samples were used for training. In model testing, $100$ samples in LJ-$001$ were used. The samples in LJ-$002$ were used for searching the optimal inference schedule.

\textbf{Conditioner}: We extract the $80$-band mel-spectrogram features with the $1024$ FFT points, $80$Hz and $8000$Hz lower and upper frequency cutoffs, and a hop length of $256$. In both training and inference process, the mel-spectrograms are computed from the ground-truth audio. As to the time embedding $t$, we follow WaveGrad \cite{WaveGrad} which replaces it with the continuous noise level $\bar{\alpha}_{t}\in (0,1)$.

\textbf{Model}: The U-Net architecture \cite{UNet} is widely used in DDPMs. We use the $15$M WaveGrad Base model \cite{WaveGrad} with the upsampling factor of $[4,4,4,2,2]$ in U-Net to match the hop length of $256$. The number of channels is $512$, $512$, $256$, $128$, $128$ respectively. A batch size of $256$ with a segmentation length of $7168$ data points is used. The optimizer is Adam \cite{Adam}.       

\textbf{Noise Schedule}: In diffusion process, the number of time steps is set to $T=1000$ and the $\beta_{t}$ linearly increases from $10^{-6}$ to $10^{-2}$. In InferGrad, for the range of $\hat{\beta}$ considered in infer loss, when $N=6$, we follow WaveGrad \cite{WaveGrad} to explore the possibilities of each step within one order of magnitude from $10^{-6}$ to $10^{-1}$. When $N=3$ and $N=2$, we explore the $3$-step schedule with $\hat{\beta}_{1} \in [10^{-6}, 10^{-4})$, $\hat{\beta}_{2} \in [10^{-4}, 10^{-2})$, $\hat{\beta}_{3} \in [10^{-1}, 1)$, and the $2$-step schedule with $\hat{\beta}_{1} \in [10^{-5}, 10^{-2})$, $\hat{\beta}_{2} \in [10^{-1}, 1)$. Some of these ranges are wider than the proposed suggestions for setting inference schedule, in order to guarantee the coverage of schedules and generalization of InferGrad. Especially, the grid search in WaveGrad \cite{WaveGrad} used a search step of $1$. In InferGrad, we use a uniform distribution for the range of each $\hat{\beta}$. Hence, we consider a continuous range rather than only consider using the step of $1$. In test, for the baseline WaveGrad, grid search is employed to find the optimal schedule. The searched optimal schedule usually follows the suggestions we presented. Hence, it is also covered in our defined ranges for InferGrad. Therefore, in evaluation, we compare the generation quality according to the optimal schedules searched for WaveGrad, instead of the ones for InferGrad. A batch of $10$ test samples with L$1$ loss of mel-scaled log-spectrum as the test metric are used for searching. 

\textbf{Infer Loss}: We define the weight $\lambda=5\times 10^{-4}$ when $N=2$ and $N=3$, and $\lambda=10^{-3}$ when $N=6$ according to the validation performance. The computation of $L_{I}$ is implemented with the publicly available auraloss \cite{auraloss}. In practice, we choose the number of resolution $M=3$, and use the FFT sizes of $[512, 1024, 2048]$ and the Hanning window lengths of $[240, 600, 1200]$.     

\textbf{Evaluations}: Considering incorporating the reverse process into training enlarges the time cost, we choose to verify our idea by fine-tuning a WaveGrad Base model \cite{WaveGrad} which is pre-trained with $1$M iterations. When $N=6$, $3$ and $2$, we continue to train this model with the objective in Equation (\ref{newtrainingobjective}) respectively. A learning rate of $5.8\times 10^{-5}$ is used \cite{PriorGrad}. Then we compare them with the baseline model at same iterations. For stable performance, we continue training with more than $250$k iterations. When comparing results given $N=6$, $3$ and $2$, for the corresponding baseline WaveGrad models, the searched optimal inference schedules are $\hat{\beta}=[0.000006, 0.00002, 0.0001, 0.001, 0.02, 0.3]$, $\hat{\beta}=[0.00005, 0.005, 0.3]$, $\hat{\beta}=[0.0001, 0.3]$ respectively. As the model performance is heavily degraded when $N=2$, the objective metric cannot effectively represent sample quality. We manually correct it as $\beta_{t}=[0.001, 0.5]$ in test. For subjective evaluation, we use both CMOS (Comparison Mean Opinion Score) and MOS (Mean Opinion Score) tests. Each CMOS result is given by $15$ judges per utterance for comparing the samples generated by the two different models. The MOS results are given by $20$ judges per utterance for evaluating the overall performance with a 5-point scale. For objective measurements, we employ log-mel spectrogram mean squared error (LS-MSE), multi-resolution STFT (MRSTFT) \cite{ParallelWaveGAN}, perceptual evaluation of speech quality (PESQ) \cite{pesq} and short-time objective intelligibility (STOI) \cite{stoi}. The PESQ and STOI are calculated with publicly available speechmetrics \footnote{https://github.com/aliutkus/speechmetrics}.

\begin{figure*}[htbp]
\centerline{\includegraphics[width=\linewidth]{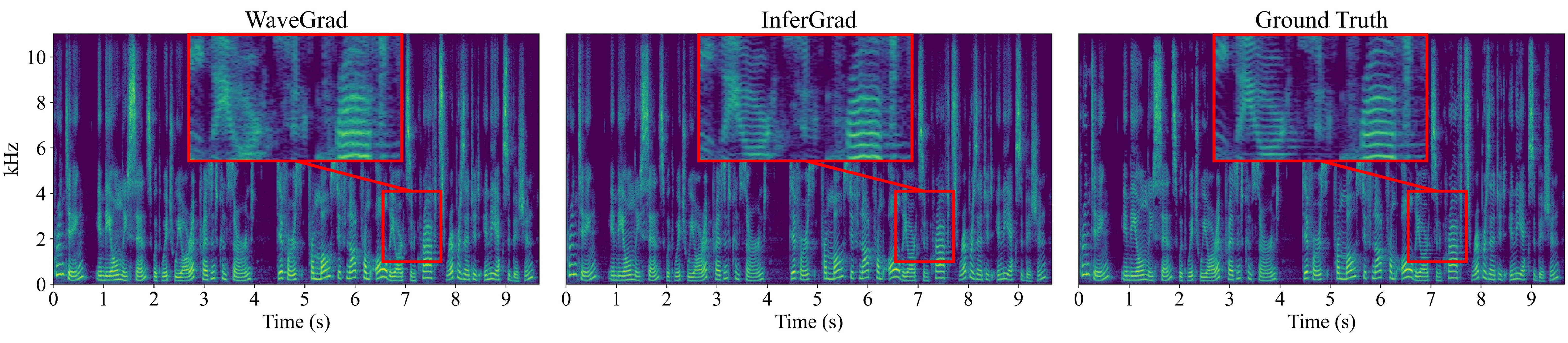}}
\vspace{-0.45cm}
\caption{The spectrum of the sample LJ$001$-$0001$ generated by WaveGrad and InferGrad with $6$ inference steps, and the ground-truth data.}
\label{fig:LJ001spectrum}
\vspace{-0.55cm}
\end{figure*}  

\vspace{-0.3cm}
\subsection{Results and Analyses}
\vspace{-0.1cm}
\subsubsection{Generation Quality}
\vspace{-0.7cm}
\begin{table}[htbp]
\centering
\caption{MOS results of InferGrad for vocoder.}{}
\begin{tabular*}{7.25cm}{@{}@{\hspace{0.25cm}}cccc@{}}
\toprule Model & \multicolumn{3}{c}{MOS (95\% confidence interval, $\uparrow$)} \\
\midrule GT & & 4.22$\pm$0.06& \\
\hline
     &N=2 &N=3 &N=6   \\
    WaveGrad &3.12$\pm$0.10 &3.43$\pm$0.09 &3.69$\pm$0.09   \\
    InferGrad &\textbf{3.69$\pm$0.08} &\textbf{3.90$\pm$0.07} &\textbf{3.97$\pm$0.07} \\
\bottomrule
\end{tabular*}
\label{MOS}
\end{table}
\vspace{-0.4cm}

\vspace{-0.45cm}
\begin{table}[htbp]
\centering
\caption{Objectives evaluations of InferGrad for vocoder.}{}
\begin{tabular*}{8.25cm}{@{}@{\hspace{0cm}}ccccc@{}}
\toprule Model& LS-MSE ($\downarrow$) & MRSTFT ($\downarrow$) & PESQ ($\uparrow$) & STOI ($\uparrow$)\\
\midrule WG2 & 0.591 & 1.569 & 2.540 & 0.927 \\
    IG2 & \textbf{0.202} & \textbf{1.238} & \textbf{3.408} & \textbf{0.967}   \\
\hline    
    WG3 & 0.429 & 1.451 & 2.901 & 0.945   \\
    IG3 & \textbf{0.121} & \textbf{1.111} & \textbf{3.506} & \textbf{0.973} \\
\hline    
    WG6 & 0.237 & 1.289 & 3.315 & 0.959 \\
    IG6 &\textbf{0.108} & \textbf{1.060} & \textbf{3.578} & \textbf{0.976}  \\
\bottomrule
\end{tabular*}
\label{objective}
\end{table}
\vspace{-0.3cm}

We report the subjective results in Table \ref{MOS} and objective results in Table \ref{objective}. As can be seen in Table \ref{MOS}, given the same inference steps when $N=6$, $3$ and $2$, InferGrad improves the sample quality by over $0.28$ MOS. Achieving similar generation quality, InferGrad accelerates inference speed by $3$x ($2$ steps for InferGrad vs $6$ steps for WaveGrad). In Table \ref{objective}, WG stands for WaveGrad, IG stands for InferGrad, IG6 stands for InferGrad with 6 steps. Table \ref{objective} shows that the objective results of both WaveGrad and InferGrad are steadily improved when $N$ increases. In comparison, for each metric, the InferGrad using $2$ iterations achieves better results than the WaveGrad using $6$ iterations. In Figure \ref{fig:LJ001spectrum}, we use the spectrum of the first test sample LJ$001$-$0001$ to demonstrate the advantages of InferGrad. For the baseline WaveGrad, we follow the publicly available implementation\footnote{https://github.com/ivanvovk/WaveGrad} where the $15$M WaveGrad Base model instead of the $23$M Large one is used for noise estimation. As this implementation is not released by the authors of WaveGrad, the baseline results may be not equal to the ones reported in WaveGrad \cite{WaveGrad}. The DiffWave \cite{DiffWave} is a similar work of WaveGrad. But the maximum number of diffusion steps is set as $T=200$ and the corresponding $\beta$ is defined as $\beta \in [10^{-4},0.02]$ in it. In comparison, the setting of $T=1000$ and $\beta \in [10^{-6},0.01]$ is used in WaveGrad. A larger number of diffusion steps $T$ and a smaller minimum noise scale $\beta_{1}$ in diffusion process may be helpful to the sample quality.

Except for considering $N=6$, $3$ or $2$ respectively, we also explore to simultaneously include them into training. With sufficient continual training iterations, e.g., $500$k, using same inference schedules, the sample quality of this model is on par with the models separately trained. In this experiment, properly defining the $\hat{\beta}$ range is important. Otherwise, the loss value may change considerably, resulting in failure of convergence. In Table \ref{generaltraining}, We use CMOS to compare the sample quality between the model fine-tuned for specified inference step (specified model for short) and the model fine-tuned for the three possibilities (general model for short): $N=6$, $3$ or $2$. As can be seen, for $2$-step and $3$-step generation, there is almost no difference between them. For $N=6$, the CMOS of general model is slightly lower than the specified model but the difference is not statistically significant. The results in Table \ref{generaltraining} demonstrate that the InferGrad is capable of considering the inference schedules with either specified iteration or a group of iterations. 

\vspace{-0.45cm}
\begin{table}[htbp]
\centering
\caption{CMOS results of the general InferGrad model}{}
\begin{tabular*}{6.5cm}{@{}@{\hspace{0.25cm}}cccc@{}}
\toprule Training objective & \multicolumn{3}{c}{CMOS ($\uparrow$)} \\
\midrule &N=2 &N=3 &N=6   \\
    Specified &0 &0 &0   \\
    General &-0.028 &0.018 &-0.067 \\
\bottomrule
\end{tabular*}
\label{generaltraining}
\end{table}
\vspace{-0.6cm}

\vspace{-0.1cm}
\subsubsection{Model Sensitivity}
\vspace{-0.1cm}
As the metric used in grid search, the L$1$ loss of mel-spectrogram features is also employed for testing the model sensitivity to inference schedule $\hat{\beta}$. Also, the mel-spectrogram feature extraction parameters follows the settings in model training. Each possibility in the proposed grid search range is tested. In Table \ref{L1loss}, we present the mean and the standard deviation (STD) of L$1$ loss. The mean represents the average quality of samples generated with each $\hat{\beta}$ in the defined search range. A large STD illustrates that the model is sensitive to $\hat{\beta}$ and it is important to search the optimal $\hat{\beta}$. As can be seen, when we reduce the number of inference steps, the WaveGrad generation quality is heavily degraded. Moreover, it becomes especially sensitive to the choice of $\hat{\beta}$. In comparison, the InferGrad keeps improving the sample quality in each setting and it is more robust to the choice of inference schedule as it considered each possibility in the range, which lowers the requirement on searching the optimal $\hat{\beta}$. 

\vspace{-0.45cm}
\begin{table}[htbp]
\setlength{\abovecaptionskip}{-0.4cm}
\setlength{\belowcaptionskip}{0.4cm}
\centering
\caption{L1 loss under different inference iterations.}{}
\begin{tabular*}{7.25cm}{@{}@{\hspace{0.25cm}}cccc@{}}
\toprule 
    Model & \multicolumn{3}{c}{L1 loss ($\downarrow$)} \\
\midrule &N=2 &N=3 &N=6   \\
    WaveGrad &1.71$\pm$0.95 &1.41$\pm$0.90 &0.41$\pm$0.08   \\
    InferGrad &\textbf{0.99$\pm$0.61} &\textbf{0.49$\pm$0.29} &\textbf{0.27$\pm$0.03} \\
\bottomrule
\end{tabular*}
\label{L1loss}
\end{table}
\vspace{-0.6cm}

\subsubsection{Ablation Study}
\vspace{-0.1cm}
We propose to use both magnitude loss and phase loss in InferGrad. Taking $2$-step generation as an example, we demonstrate the effectiveness of phase information by removing it from infer loss. We tested the sample quality and illustrate the result with CMOS score in Table \ref{ablation}. In the result, when we remove the phase loss from training objective, with the optimization of magnitude loss, there exist artifacts in synthesized samples because the magnitude loss cannot fit human evaluations perfectly. By using magnitude and phase together, we distinctively reduced the gap between objective measurement and human perception. Moreover, we didn't remove the magnitude information from infer loss because in that way important information will be discarded rendering the distance not meaningful. 

\vspace{-0.4cm}
\begin{table}[htbp]
\centering
\caption{CMOS results of ablation study for InferGrad}{}
\begin{tabular*}{4.5cm}{@{}@{\hspace{0.225cm}}cc@{}}
\toprule 
Loss Function & {CMOS ($\uparrow)$} \\
\midrule 
InferGrad & 0   \\
InferGrad - $L_{pha}$ & -0.140   \\
InferGrad - $L_{mag}$ & - \\
\bottomrule
\end{tabular*}
\label{ablation}
\end{table}
\vspace{-0.35cm}

\section{Conclusions and Discussions}
\vspace{-0.1cm}
By considering the inference process in training, the proposed InferGrad approach has been shown to improve DDPMs generation quality when the number of inference steps is small, which enables fast and high-quality sampling. Moreover, InferGrad effectively enhances the robustness of DDPMs to inference schedule. This paper is a preliminary study on the feasibility of incorporating fast sampling process into training. In the future, several aspects are worthy of exploring. First, other sample quality evaluation metrics, e.g., the distance in latent space, can be employed in the infer loss of InferGrad to minimize the gap between the objective measurement and human evaluation. Second, except for inference schedule, the training noise schedule should be carefully selected given sampling with a small number of steps. Third, the network architecture can be improved to enhance the noise estimation ability.  
 

\vfill\pagebreak

\bibliographystyle{IEEEbib}
\bibliography{mybib}

\end{document}